\documentclass[prl,twocolumn,superscriptaddress]{revtex4}
\usepackage{amsmath}
\usepackage{amssymb}
\usepackage{graphicx}
\usepackage{subfigure}
\usepackage{verbatim}
\usepackage{hyperref}
\usepackage{xfrac}
\usepackage{amsfonts}
\usepackage{bbm}
\usepackage[normalem]{ulem}
\usepackage{color}

\newcommand{\beginsupplement}{%
        \setcounter{table}{0}
        \renewcommand{\thetable}{S\arabic{table}}%
        \setcounter{figure}{0}
        \renewcommand{\thefigure}{S\arabic{figure}}%
        \setcounter{equation}{0}
        \renewcommand{\theequation}{S\arabic{equation}}%
      }

\makeatletter
\newcommand*{\rom}[1]{\expandafter\@slowromancap\romannumeral #1@}
\makeatother

\begin{document}
\title{Instabilities and Solitons in Minimal Strips}
\author{Thomas Machon}
\author{Gareth P. Alexander}
\affiliation{Department of Physics and Centre for Complexity Science, University of Warwick, CV4 7AL, United Kingdom.}
\author{Raymond E. Goldstein}
\author{Adriana I. Pesci}
\affiliation{Department of Applied Mathematics and Theoretical Physics, Centre for Mathematical Sciences, University of Cambridge, Wilberforce Road, Cambridge CB3 0WA, United Kingdom.}
\begin{abstract}
We show that highly twisted minimal strips can undergo a non-singular transition, unlike the singular transitions seen in the M\"obius strip and the catenoid. If the strip is non-orientable this transition is topologically frustrated, and the resulting surface contains a helical defect. Through a controlled analytic approximation the system can be mapped onto a scalar $\phi^4$ theory on a non-orientable line bundle over the circle, where the defect becomes a topologically protected kink soliton or domain wall, thus establishing their existence in minimal surfaces. Experimental studies of soap films confirm these results and demonstrate how the position of the defect can be controlled through boundary deformation. 
\end{abstract}
\maketitle

Minimal surfaces, critical points of the area functional, are geometric motifs that appear across physics. From their early identification in soap films~\cite{Plateau}, they have since been identified in a variety of places in condensed matter physics~\cite{schoen12} including as the boundary in bicontinuous phases~\cite{longley83,hyde84}, smectic liquid crystals~\cite{kamien99,matsumoto11} as well as in other areas such as monopoles and twistor theory~\cite{hitchin82,hitchin87}. The simplest minimal surface is the plane, which is also the only stable complete minimal surface~\cite{fischercolbrie80}, and thus any other soap film surface must become unstable if the boundary is deformed beyond a critical conformation. Studies of these instabilities in soap film annuli have typically focused on the catenoid~\cite{cryer92,chen97} and, more recently, the M\"{o}bius strip~\cite{goldstein10, goldstein14, pesci15}. In both of these cases, as the boundary wire is varied one sees an instability that leads to a singularity and subsequent topological change in the surface -- to two discs in the case of the catenoid and a single disc in the case of the M\"{o}bius strip. More generally, these instabilities illustrate how surface tension serves to control morphology, and morphological changes, in a simple but generic system. 

Here we discuss instabilities in highly twisted minimal strips. We demonstrate that the collapsing instabilities that one sees for the catenoid and M\"{o}bius strip do not extend to an arbitrary number of twists. Instead, one observes buckling instabilities that lead to non-singular transitions in the surface, similar in nature to the non-singular `headphone' transitions studied by Courant~\cite{courant40}, and the creation of ribbon-like structures, analogous to those seen in the helicoid~\cite{boudaoud99}. In geometric terms, the change in the morphology of the strip is from twist to writhe. We show that if the strip is non-orientable the instability is topologically frustrated and the resulting ribbon contains a topological defect or domain wall, which we show in a series of experiments, demonstrating the topological robustness of the effect. The simplest examples of domain walls are found in one-dimensional systems with two groundstates such as polyacetylene~\cite{su79,su80,heeger88} and its mechanical analogues~\cite{kane13,chen14}, where the continuum limit is a $\phi^4$ field theory~\cite{MantonSutcliffe,Vachaspati}. By mapping the structure of the non-orientable strip onto such a theory, we demonstrate the existence of topological solitons in minimal surfaces, where the domain wall becomes a $\mathbb{Z}_2$ kink soliton, topologically protected by the non-orientability of the surface. Finally, we show how the location and motion of the defect can be controlled through manipulation of the boundary curve.
\begin{figure}[t]
\includegraphics{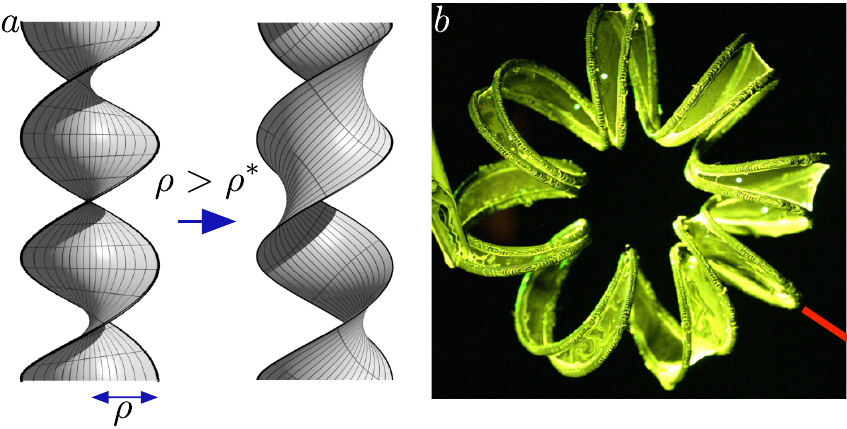}
\caption{(a): The instability in the helicoid. For $\rho < \rho^\ast \approx 1.509$ the helicoid is stable, but above this critical value there is an an instability which leads to the formation of a ribbon, shown on the right. (b): Photograph of a soap film on a circular helix frame containing a kink soliton, indicated by a red line. The radius of the helix is sufficiently large that a ribbon is preferred over a circular helicoid, however the boundary curve has an odd number half twists making the surface non-orientable and leading to the presence of the defect. The defect indicates where the surface is locally helicoidal and can be thought of as interpolating between two ribbons that are $\pi$ out of phase with each other.} 
\label{fig:2}
\end{figure}

The instabilities of a soap film are found by studying the second variation of area, since the energy derives entirely from surface tension, which leads to the Schr\"{o}dinger operator with Dirichlet boundary conditions~\cite{Neel} 
 \begin{equation}
 -\nabla^2+2K,
 \label{eq:stab_op}
 \end{equation}
where $K$ is the Gaussian curvature of the surface. The eigenfunctions of this operator correspond to the normal modes of a soap film realising the particular surface, with the squared frequency, $\omega^2$, given by the eigenvalues as $\lambda_i = d h \omega_i^2 / \sigma$ where $h$, $d$ and $\sigma$ are the thickness, density and surface tension of the soap film respectively~\cite{boudaoud99}. For a physical soap film spanning a wire frame, instability occurs if the motion of the frame pushes the lowest eigenvalue of \eqref{eq:stab_op} below zero. It follows that an observed instability of a soap film is both a solution of $(-\nabla^2+2K) \psi =0$, known as a Jacobi field, and the groundstate of \eqref{eq:stab_op}. 

\begin{figure}
\includegraphics{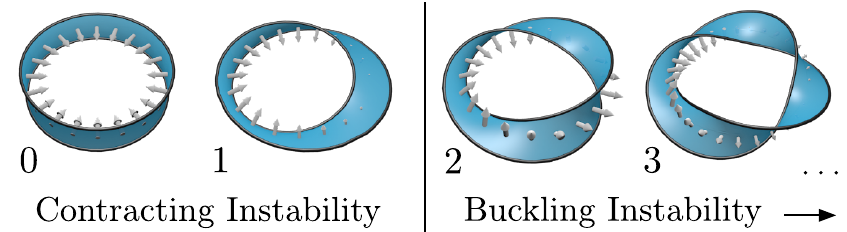}
\caption{Bent helicoids with $q=0$, 1, 2 and 3 half twists. The arrows indicate the nature of the instability. For $|q|>1$, the instability leads to a buckling transformation, rather than the collapse that occurs for $q=0,1$.}
\label{fig:jac}
\end{figure}

$\psi$ represents an infinitesimal deformation of the surface in the normal direction; the mode of deformation at the point of instability. Its nature depends upon the topological type of the minimal strip, determined by an integer, $q$, equal to the number of half twists in the strip; the catenoid corresponds to $q=0$ and the M\"{o}bius strip to $q = \pm 1$. Examples of surfaces with arbitrary $q$ are given by the bent helicoids~\cite{meeks07,mira06}, shown in Fig.~\ref{fig:jac}, which are determined by their Weierstrass data, $g(w) = -w (w^{q/2}+i)/(i w^{q/2}+1)$ and $d f = (2w)^{-1}(w^{q/2}+w^{-q/2}) dw$, and contain the unit circle in the $x$-$y$ plane. If $q$ is odd, the surface is non-orientable, and orientable is $q$ is even. They have $q$-fold rotational symmetry about their axial direction and the geometry of the surfaces is pure twist with no writhe. From Bloch's theorem we know that, in terms of conformal coordinates $w = 2(u + iv)/q$ where $u \in [0, q \pi)$ runs along the strip, $\psi$ is of the form $\psi(u,v) = \psi_q(u,v)$ for an orientable strip, where $\psi_q(u,v) = \psi_q(u+ \pi,v)$. If the surface is non-orientable, then the surface normal, $\mathbf{n}$, reverses sign upon traversal of the strip, meaning that $\psi$ necessarily contains a nodal line. By passing to the double cover and considering the first excited state, one may observe that $\psi$ must be of the form $\psi(u,v) = \sin(u/q) \chi_q(u,v)$ in the non-orientable case, where $\chi_q$ is a $q$-fold periodic function as in the orientable case, and so contains a nodal line which marks the position of the solition (here $u=0$).

\begin{figure}
\includegraphics{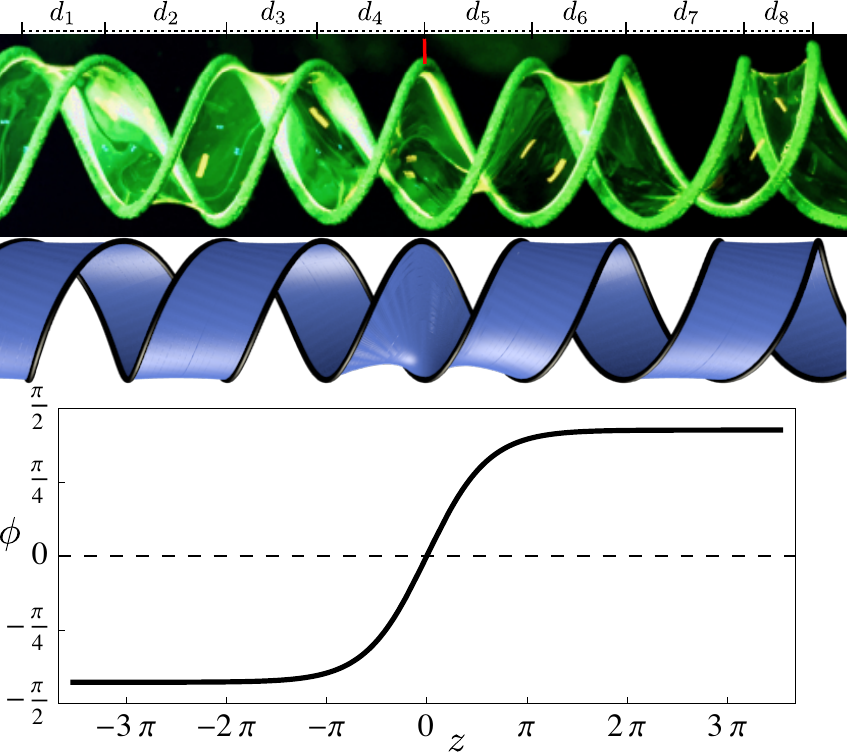}
\caption{Comparison of the tanh solution for the ruled approximation to and experimental realisation of the soliton spanning a segment of helical boundary curves. \textit{Bottom}: graph of the tanh solution. \textit{Middle}: Plot of the ruled surface with similar $\rho$ to the experiment, with $\phi = m \tanh \big ( \sqrt{2 \lambda} m z \big )$. \textit{Top}: Experimental image of the soliton, marked with a red line, with helical boundary curves. Note that to the left of the soliton $d_{2n} > d_{2n \pm 1}$, whereas to the right $d_{2n} < d_{2n \pm 1}$, indicating a deformation of the boundary.}
\label{fig:1}
\end{figure}

While the functions $\psi_q$ and $\chi_q$ have $q$-fold symmetry, because of the twisted nature of the circular helicoids $\mathbf{n}$ has only $q/2$-fold rotational symmetry and reverses sign under a rotation of $2 \pi/q$, that is $\mathbf{n}(u,v) = -\mathbf{n}(u + \pi, v) = \mathbf{n}(u + 2\pi, v)$. Consequently, for $|q|>1$ the instability pushes adjacent segments of the surface in opposite directions, as shown in Fig.~\ref{fig:jac}, which leads to a buckling of the surface to form a ribbon, illustrated in Fig.~\ref{fig:2}, rather than the collapse that characterises both the catenoid ($q=0$) and M\"obius strip ($|q|=1$). In the non-orientable case, the nodal line in $\psi$ means that the instability is frustrated. As such, when the non-orientable strip undergoes its instability there is a helicoidal defect created in the ribbon surface, which one can think of as a local interpolation between two ribbon surfaces. This defect marks the location of a topological soliton in the soap film. The topological nature of the phenomenon means that it is robust to perturbations and deformations of the surface, or indeed the exact shape of the bounding frame: it is a generic feature of non-orientable minimal strips with multiple half twists ($|q|>1$). We have realised examples experimentally on circular frames, shown in Fig.~\ref{fig:2}(b), and also on frames resembling the shape of an athletics stadium, shown in Fig.~\ref{fig:1} (straight portion only). The latter allows the defect to be studied in the straight region of the frame where the boundary is an ordinary straight double helix. This setting simplifies some of the analysis without sacrificing any features.

A prototypical local model for this buckling instability is the helicoid, discussed by Boudaoud {\sl et al.}~\cite{boudaoud99}. The surface is given by $\Sigma(u,v)=(\sinh v \cos u, \sinh v \sin u, u)$, where $u \in \mathbb{R}$ and $\sinh v \in [-\rho, \rho]$. As $\rho$ grows beyond a critical threshold, $\rho^\ast$, this helicoid undergoes the aforementioned buckling instability to form a ribbon, shown in Fig.~\ref{fig:2}(a). The instability can result in one of two separate ribbons, depending on the sign of the initial instability~\cite{boudaoud99}, which are related to each other by a translation of half a period in the vertical direction. An explicit form for the Jacobi field is given by the normal deformation $\mathbf{n} \cdot d \Sigma_t / dt$ of a one-parameter family of minimal surfaces, $\Sigma_t$~\cite{meeks11,note_jac}. For the helicoid, one uses the Bonnet transformation  
\begin{equation}
\Sigma_t(z) = \textrm{Re}\big [ e^{i t} (\cos z, \sin z, -i z )\big],
\end{equation}
which corresponds to the helicoid at $t=0$ and the catenoid at $t= \pi/2$. This gives $\psi(v) = 1- v \tanh v$ and the critical threshold of $\rho^\ast \approx 1.509$, given by $\psi(\textrm{asinh}(\rho^\ast)) =0$. We note here that for the catenoid and the M\"{o}bius strip, one can obtain the relevant Jacobi field using a scale transformation, $\Sigma_t=(1-t)\Sigma$, as the instability corresponds to a shrinking of the surface. 

We now consider the case of a twisted strip containing $q$ half twists, bounded by a double helix frame of radius $\rho$ in a periodic domain $(x,y,z) \sim (x,y,z+ \pi q )$. As before, for $\rho$ small, the minimum area solution is a helicoid, and as $\rho$ grows above a critical value the strip becomes unstable. If $q$ is even then the strip is orientable and the Jacobi field for the instability is identical to that for the helicoid. If $q$ is odd the surface is non-orientable and $\psi(u,v)$ must satisfy $\psi(u,v)=- \psi(u + q \pi, - v)$. The simple structure of the helicoid allows for the Jacobi field in the non-orientable case to to be solved for exactly  (see Supplementary Information) and an explicit form can be given as 
\begin{multline}
\psi(u,v) = \\ \big ( Q_2^{1/q}(0) P_1^{1/q}(\nu) - P_2^{1/q}(0) Q_1^{1/q}(\nu) \big)  \sin(u/q),
\label{eq:non_jac}
\end{multline}
where $P_n^m(x)$ and $Q_n^m(x)$ are the Legendre functions of the first and second kind respectively and $\nu =\tanh v$. The critical value $\rho^\ast_q$ now depends on $q$ and forms a decreasing sequence with limiting value $\rho^\ast$. Hence, for any finite $q$, $\rho^\ast_q > \rho^\ast$, and the corresponding non-orientable helicoid enjoys a slightly greater degree of stability than its orientable brethren.

\begin{figure}
\includegraphics{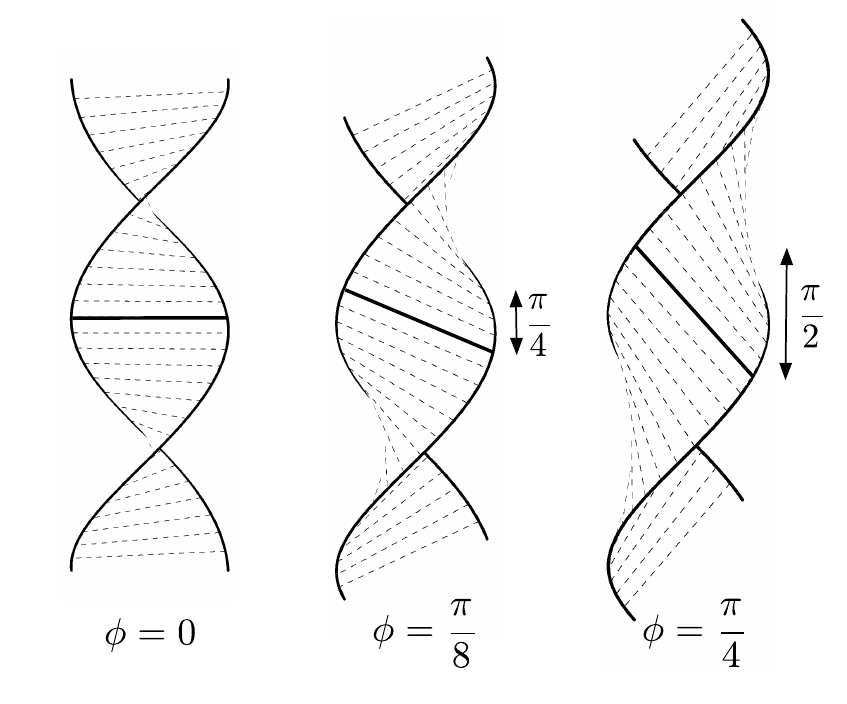}
\caption{The ruled approximation used to find the shape of the kink soliton. The surface consists of straight lines connecting points on one boundary helix with a phase difference of $2\phi$ on the opposite boundary helix. The surface is only minimal for $\phi=0$, which gives the helicoid, increasing $|\phi|$ decreases the quality of the approximation.}
\label{fig:ruled}
\end{figure}

In the orientable case exact forms can be given for the resulting ribbon surfaces~\cite{boudaoud99}, (including with a circular axis~\cite{meeks07,mira06, note_bent_helicoid}). However, to find the form of the ribbon surface containing the topological defect we turn to an approximation. The helicoid is well-known to be the only ruled minimal surface, and for $\rho$ just above the critical value, the resulting ribbon is close in form to the helicoid. We therefore approximate the system by a set of ruled surfaces given by the formula
\begin{equation}
\Sigma(z,r) = r c_1(z-\phi(z))+ (1-r)c_2(z+\phi(z)),
\label{eq:ruled}
\end{equation}
where $r \in [0,1]$, $z \in [0,q \pi)$, and the boundary curves are given by $c_1(z)=(\rho \cos z, \rho \sin z, z)$ and $c_2(z) = (-\rho \cos z, -\rho \sin z, z)$. The surface given by \eqref{eq:ruled} connects points on one helical boundary to the other by straight lines, offset by a phase of $2\phi(z)$, as illustrated by Fig.~\ref{fig:ruled}. $\phi=0$ gives (exactly) the helicoid, while a constant value of $\phi \neq 0$ gives a ribbon-like surface. To test the validity of this approximation, one can solve to find the value of $\phi$ that minimises the area functional 
\begin{equation}
A= \int_0^{q \pi} \int_0^1 |\partial_z \Sigma \times \partial_r \Sigma | \; dr dz ,
\label{eq:area}
\end{equation}
for a given boundary radius $\rho$ (see Supplementary Information). For small $\rho$, $\phi=0$ gives the minimal solution as a helicoid. As in the general case there is a pitchfork bifurcation at a finite value, $\bar{\rho}$, above which there are two non-zero equilibrium values of $\phi$, $\pm \bar{\phi}(\rho)$. As $\rho$ grows large $ \bar{\phi} \to \pi/2$, corresponding to a ribbon surface lying on the surface of a cylinder of radius $\rho$. The transition radius $\bar{\rho}  \approx 1.511$ is close to the value for the instability of the helicoid of $\rho^\ast \approx 1.509$ though necessarily slightly greater.

To study the shape of the soliton on the periodic helicoid, we must allow $\phi$ to vary. Just above the transition radius, $\bar{\rho}$, the equilibrium values of $\phi$, $\pm \bar{\phi}$, are close to zero. As such we expand to low order in $\phi$, and performing the $r$ integral in \eqref{eq:area} we find 
\begin{equation}
A \approx \int \alpha + \beta \phi^2 + \gamma \phi^4 + \delta \big(\phi^\prime \big )^2 dz,
\label{eq:lag}
\end{equation}
where the values of $\alpha$, $\beta$, $\gamma$ and $\delta$ depend on $\rho$ (for explicit forms please see the Supplementary Information) and we have temporarily suppressed the limits on the integral. \eqref{eq:lag} gives the spatial part of the Lagrangian for a scalar $\phi^4$ theory on the circle~\cite{MantonSutcliffe,Vachaspati}. The helical defect, shown in Figs.~\ref{fig:2} and \ref{fig:1}, therefore corresponds to a kink soliton in the $\phi^4$ theory in \eqref{eq:lag}.

In the periodic domain, the global topology in the non-orientable case is enforced by demanding that $\phi(z) = -\phi(z+q \pi)$, and the existence of the kink is topologically protected. As a consequence of the non-orienability, the state described by $\phi$ is equivalent to that described by $-\phi$ and the kink is equivalent to the anti-kink. In general, the topological classification of a scalar field on a circle is given by a $\mathbb{Z}_2$ invariant, the first Stiefel-Whitney class, $w_1 \in H^1(S^1; \mathbb{Z}_2)$, of the line bundle associated to $\phi$, and counts the number of solitons in the system modulo 2; the non-orientable case then corresponds to $w_1=1$.

\eqref{eq:lag} can be solved exactly to give a solution in terms of elliptic functions as 
\begin{equation}
\phi  = \pm \sqrt{\frac{4 m^2 \lambda - b^2 }{2 \lambda}}  \;\textrm{sn} \left( b z, \frac{4 m^2 \lambda}{b^2}-1\right),
\label{eq:sol}
\end{equation}
where $m^2 = -\beta /(2 \lambda)$, $\lambda = \gamma / (2\delta)$ and $b$ is a constant. For an $n$ ($n$ odd) kink solution the value of the constant $b$ is found by satisfying the boundary conditions $\phi(z) = -\phi(z+q \pi)$, which can be written as
\begin{equation}
\frac{ q \pi}{n} =  \frac{2}{b}K \left( \frac{4 m^2 \lambda}{b^2}-1 \right),
\end{equation}
where $K(x)$ is the complete elliptic integral of the first kind. We will focus on the lowest area case where $n=1$.
 
The system given by \eqref{eq:sol} has two length-scales, the length of the circle, $q \pi$, and the preferred width of the soliton, $1/\sqrt{2 \lambda} m$. If $q \pi  \sqrt{2 \lambda} m \gg 1$, which can be satisfied by having a large number of twists in the helical boundary ($q$ large), then one can well-approximate the solutions on the circle with the standard kink solution, which is simply given by
\begin{equation}
\phi(z) = \pm m \tanh \big ( \sqrt{2 \lambda} m z \big ).
\label{eq:stat}
\end{equation}
Note that in the vicinity of $\bar{\rho}$,  $\sqrt{2 \lambda} m  \sim (\rho-\bar{\rho})^{1/2}$ and so this approximation requires, for finite $q$, that $\rho$ is sufficiently above the critical value. Fig.~\ref{fig:1} shows a comparison of the solution obtained in this approximation with an experimental image, to which it shows a good degree of similarity. 

The width of the soliton has a limit of $\sqrt{2}$ as $\rho \to \infty$. While this is outside the range of applicability of our approximations, from a topological perspective one can show that a finite limiting width must exist. The existence of the defect implies that the curve $x=y=0$, and any isotopic curve lying in the interior of the helical frame, must intersect the surface once. As such, when projected onto the $z=0$ plane the soliton must span a disc of radius $\rho$. The smallest $z$ interval within which the boundary of this disc can be mapped onto the boundary of the helicoid is $\pi$, giving a lower bound for soliton width. Using surface evolver~\cite{brakke92} surfaces attaining close to this limit can be found for $\rho \gtrapprox 3$. We note that a fuller discussion of the equilibrium shape of the experimental system would be to consider a competition between the surface tension of the film and the elastic energy of the boundary wire, the so-called Euler-Plateau problem~\cite{giomi12}. 

The Lagrangian \eqref{eq:lag} has a continuous translational symmetry, corresponding to the continuous screw symmetry of the helicoidal frame. This leads to the presence of a Goldstone mode~\cite{Vachaspati}, localised to the defect, and ultimately allows it to move. In physical realisations of these systems, the boundary curve typically does not possess this symmetry, and the location and motion of the solition is driven by the inhomogeneities in the boundary. A simple example of this is Fig.~\ref{fig:2}(b), and more generally the bent helicoids, where the axis is circular. In this case the continuous screw symmetry is broken to a discrete $q$-fold rotational symmetry, leading to $q$  preferred locations for the defect.  Experimental, numerical and theoretical~\cite{note_surf} results all indicate that on a circular frame the defect is located such that the line in the centre of the defect points towards the centre of the circle, as shown in Fig.~\ref{fig:2}(b).

Making the axis circular allows one to specify where the defect will be up to the $q$-fold degeneracy, but does not allows for easy manipulation of the defect position. To control the defect's location, and induce motion, one can construct simple deformations of a straight helicoid. Moving the helicoid along its local axis in the periodic domain can be achieved by setting the boundary curves to $\big (\rho \cos z, \rho \sin z, z + g(z)+h(z) \big )$ and $\big (-\rho \cos z, -\rho \sin z, z + g(z)-h(z) \big)$, where $z \in [0, q \pi)$ and the shifts are determined by the functions $g$ and $h$. In the non-orientable case ($q$ odd) continuity demands that $g(z + q \pi) = g(z)$ and $h(z+q \pi) = -h(z)$. $g(z)$ controls the local pitch of the helicoid, given by $1+ g^\prime$, whereas $h(z)$ controls the local phase difference between points on boundary curves of equal height. A deformation with $h(z)$ constant and $g(z)=0$ does not distort the boundary curves, and thus costs no elastic energy locally. For an orientable helicoid, $h(z)$ controls which of the two possible ribbons are energetically favourable; in the ruled approximation, setting $h$ leads to linear and cubic terms in \eqref{eq:lag} such that for $h>0$, $\phi<0$ is favoured over $\phi>0$ and vice-versa. In the non-orientable case, $h$ must satisfy $h(z) = - h(z+ q \pi)$, so cannot be constant everywhere and must contain an odd number of zeros. These zeros represent places where the preferred ribbon type changes from one to another, and are consequently favoured places for defects to be located. Fig.~\ref{fig:1} shows an experimental image of a defect on a straight helical frame. If $d_i$ denotes the distance between two points on the boundary curves of the same phase, as indicated in Fig.~\ref{fig:1} then $d_i = \pi \pm \big ( h(\tilde{z})+h(\tilde{z}+\pi) \big)$ with the sign alternating between plus and minus on consecutive intervals and where $\tilde{z}$ is the value of $z$ on the first measuring point. From the experimental image, one can see that to the left of the soliton $d_{2n} > d_{2n \pm 1}$, whereas to the right $d_{2n} < d_{2n \pm 1}$, indicating that the sign of $h$ has changed and the defect is located at a zero of the function $h$. In this way, by controlling the shape of the boundary through the function $h$, one can control the location of the defect. In the supplementary material we show videos of boundary deformations creating motion of the soliton through this mechanism. 

\acknowledgements{
We would like to thank D. Michieletto, D. Papavassiliou, M. Contino,  J.A. Cohen, M. Polin, V. Kantsler, M.S. Turner, G. Rowlands, H.K. Moffatt and R.B. Kusner for useful discussions. This work was supported in part by the UK EPSRC through Grant No. A.MACX.0002 (TM and GPA) and an EPSRC Established Career Fellowship (R.E.G. and A.I.P.). TM also supported by a University of Warwick Chancellor's International Scholarship and by a University of Warwick IAS Early Career Fellowship. 
}

\onecolumngrid
\pagebreak
\beginsupplement

\section*{Supplementary Information}

\subsection{I. Instability of the Periodic Helicoid}

Given the standard conformal parameterisation of the helicoid, $\Sigma : \mathbb{R}^2 \to \mathbb{R}^3$, as:
\begin{equation}
\Sigma(u,v) = \bigl( \cos u \sinh v, \sin u \sinh v, u \bigr)
\end{equation}
one finds that the associated Jacobi equation, pulled back to $\mathbb{R}^2$, is simply
\begin{equation}
\nabla^2 \psi(u,v) +  2\, \textrm{sech}^2 v \; \psi(u,v) =0.
\end{equation}
We now assume that the helicoid is in a periodic domain given by the relationship $(x,y,z) \sim (x,y,z+ \pi q )$.  These symmetries dictate that the Jacobi field should satisfy the periodicity condition 
\begin{equation}
\psi(u,v)=(-1)^q \psi(u + q \pi, (-1)^q v).
\end{equation}
If $q$ is even, the strip is orientable and the analysis is identical to the standard helicoid. When $q$ is odd, this is no longer the case and we assume a solution of the form $\psi (u,v) = \sin(u/q) f(v)$ with $f(v)=f(-v)$, which requires us to solve the 1-d problem
\begin{equation}
\frac{d^2 f}{d v^2} + 2\, \textrm{sech}^2 v f(v) = \frac{1}{q^2} f(v),
\end{equation}
where $q$ is odd. This equation has the general solution
\begin{equation}
f(v) =  A P_1^{1/q}(\tanh v) + B Q_1^{1/q}(\tanh v),
\end{equation}
where $P_n^m(x)$ and $Q_n^m(x)$ are the Legendre functions of the first and second kind respectively. The symmetry $f(v)=f(-v)$ is achieved by setting $A = Q_2^{1/q}(0)$ and $B=-P_2^{1/q}(0)$. The value of the critical radius, $\rho_q^*$, at which the instability occurs is given by $f(\textrm{asinh}(\rho_q^*))=0$. 

\subsection{II. The Area of Ruled Surfaces}

To understand the lowest area solutions for the ruled approximation, we first write down the energy of our system, simply given by the area multiplied by the surface tension
\begin{equation}
E = \sigma \int_\Sigma d A.
\end{equation}
The surface $\Sigma$ is given by the family of straight lines
\begin{equation}
\Sigma(z,r) = r c_1(z-\phi(z))+ (1-r)c_2(z+\phi(z)),
\label{eq:ruledS}
\end{equation}
connecting points on the helical boundary curves $c_1(z)=(\rho \cos z, \rho \sin z, z)$ and $c_2(z) = (-\rho \cos z, -\rho \sin z, z)$, where $r \in [0,1]$, $z \in [0,q \pi)$. 
Now to compute the area of the surface 
\begin{equation}
A = \int_{0}^{2 \pi} \int_0^1 \sqrt{\det g} \; d r d z,
\end{equation}
we need to evaluate the determinant of the metric, given as
\begin{equation}
\sqrt {\det g} = | \partial_z \Sigma \times \partial_r \Sigma|.
\end{equation}
Explicitly we have for constant $\phi$
\begin{equation}
\det g = \frac{1}{2} \rho^2 \bigg ( 4 \cos^2 \phi \big ( 2+ (1-2r)^2 \rho^2 + (1-2r)^2 \rho^2 \cos 2 \phi \big) + 8 \phi \big ( \sin 2 \phi + \phi [ 1+2r(r-1)(1+\cos 2 \phi) ] \big) \bigg) .
\end{equation}
To evaluate the area we need to compute the integral of $\sqrt{\textrm{det}\; g}$ over $r$. We can write the integrand as the square root of a polynomial
\begin{equation}
\sqrt{\det g} = \sqrt{a r^2 + br + c} = \sqrt{a}\sqrt{(r+b_1)^2+c_1}
\end{equation}
which has the indefinite integral 
\begin{equation}
I(r) =\frac{\sqrt{a}}{2}(b_1+r)\sqrt{\det g} + \frac{\sqrt{a} c_1}{2} \ln \left (b_1+r + \sqrt{\det g}\right),
\end{equation}
and so since $\phi$ is constant $E =\sigma q \pi (I(1)-I(0))$, which is an expression involving $\rho$ and and $\phi$, which can be minimised.

Allowing $\phi$ to vary means that we can no longer do the $z$ integral, instead we take a limit to find the one-dimensional system
\begin{equation}
\int \alpha + \beta \phi^2 + \gamma \phi^4 + \delta \big(\phi^\prime \big )^2 dz,
\end{equation}
where
\begin{gather}
\alpha = \rho \sqrt{1+ \rho^2} + \textrm{asinh} \rho \\
\beta = \frac{\rho\sqrt{1+ \rho^2}(1- 2 \rho^2)+(4 \rho^2-1 )\textrm{asinh} \rho}{2 \rho^2} \\
\gamma = \frac{1}{24 \rho^4(1+\rho^2)} \bigg (\rho \sqrt{1+\rho^2}(8 \rho^6-46\rho^4 + 33 \rho^2-9) + (40 \rho^6+4 \rho^4-27 \rho^2+9) \textrm{asinh} \rho \bigg) \\
\delta = \frac{\rho \sqrt{1+ \rho^2}(3+ 2 \rho^2)- (3+4 \rho^2)\textrm{asinh} \rho}{2 \rho^2}
\end{gather}
Note that the critical value $\bar{\rho}$ is such that $\beta(\bar{\rho})=0$. Note also that $ \pi q\alpha$ gives the area of the helicoid, which corresponds to $\phi=0$. 

\subsection{III. Videos of Soliton Motion}

Videos are of a soap solution made with 1 part Fairy washing-up liquid, 1 part glycerol and 2 parts water. Soap films are illuminated under blue LEDs via addition of fluorescein. Frames were manufactured using 3d printing with PLA (first video) and PA 2200 (second video).

\vspace{5mm}

The first video shows motion of the soliton on a circular frame with diameter $9.4$cm, height $3.4$cm and $q=15$, corresponding to $\rho\approx 1.7$. The frame is controlled manually at two points causing the soliton to move in a counter-clockwise direction from the upper left part of the frame to the lower right, converting a region of wide ribbon into one of narrow ribbon. It stops moving when it has fully converted the ribbon. Video shot using a Phantom V641 high speed camera with a Zeiss Makro-Planar 50mm f/2.0 lens at 300 fps. 

\vspace{5mm}

The second video shows motion of the soliton on a straight portion of a `hippodrome' frame in the shape of an athletics stadium. The frame is controlled manually at two points at the edge of the field of view, allowing defects to be created in the straight region and moved through it by gentle deformation. The total frame dimensions are $15.76\times 7.23 \times 1.54$cm with a helical radius of $0.75$cm, giving $\rho \approx 3.1$  and $q=41$. Video shot at 240 fps with an iPhone 6 camera. 


\begin{thebibliography}{99}
\bibitem{note_jac} This follows since \eqref{eq:stab_op} is the second variation of area. Jacobi fields can therefore be found as the tangent vectors to one-parameter families of critical points of the area functional.
\bibitem{note_bent_helicoid} The ribbon surfaces on a circle can be obtained through a Bonnet transformation of the (orientable) bent helicoids.
\bibitem{note_surf} This can be done by expanding in the reciprocal radius of the axis circle, and treating the resulting modified one dimensional Lagrangian with perturbation theory.
\end{thebibliography}
\end{document}